\documentclass[aps,preprint,showpacs,amsmath,amssymb,floatfix,pre]{revtex4}

\usepackage{graphicx}
\usepackage{dcolumn}
\usepackage{bm}

\begin{document}
\draft

\title{Molecular Dynamics Study\\ of Orientational Cooperativity in Water}

\author{Pradeep Kumar,$^1$ Giancarlo Franzese,$^2$ Sergey
  V. Buldyrev,$^{1,3}$ and H. Eugene Stanley$^1$} 

\address{$^1$Center for Polymer Studies and Department of Physics\\ Boston
  University, Boston, MA 02215 USA\\ $^2$ Departament de F\'{\i}sica
  Fonamental, Universitat de Barcelona, Diagonal 647, Barcelona 08028,
  Spain\\ $^3$Department of Physics, Yeshiva University\\ 500 West 185th
  Street, New York, NY 10033 USA}

\date{kfbs.tex ~~~ 21 October 2005}

\begin{abstract}

Recent experiments on liquid water show collective dipole orientation
fluctuations dramatically slower then expected (with relaxation time $>$ 50
ns) [D. P. Shelton, Phys. Rev. B {\bf 72}, 020201(R) (2005)].  Molecular
dynamics simulations of SPC/E water show large vortex-like structure of
dipole field at ambient conditions surviving over 300 ps [J. Higo at
al. PNAS, {\bf 98} 5961 (2001)]. Both results disagree with previous results
on water dipoles in similar conditions, for which autocorrelation times are a
few ps. Motivated by these recent results, we study the water dipole
reorientation using molecular dynamics simulations in bulk SPC/E water for
temperatures ranging from ambient 300 K down to the deep supercooled region
of the phase diagram at 210 K. First, we calculate the dipole autocorrelation
function and find that our simulations are well-described by a stretched
exponential decay, from which we calculate the {\it orientational
autocorrelation time} $\tau_{a}$. Second, we define a second characteristic
time, namely the time required for the randomization of molecular dipole
orientation, the {\it self-dipole randomization time} $\tau_{r}$, which is an
upper limit on $\tau_{a}$; we find that $\tau_{r}\approx 5 \tau_{a}$. Third,
to check if there are correlated domains of dipoles in water which have large
relaxation times compared to the individual dipoles, we calculate the
randomization time $\tau_{\rm box}$ of the site-dipole field, the net dipole
moment formed by a set of molecules belonging to a box of edge $L_{\rm
box}$. We find that the {\it site-dipole randomization time} $\tau_{\rm
box}\approx 2.5 \tau_{a}$ for $L_{\rm box}\approx 3$\AA, i.e. it is shorter
than the same quantity calculated for the self-dipole. Finally, we find that
the orientational correlation length is short even at low $T$.

\end{abstract}

\maketitle

\section{Introduction}
Cooperative motion of water molecules \cite{review} has been widely
investigated in recent years, both by experiments \cite{experiment,experiment2,experiment3,experiment4,experiment5,experiment6,experiment7,experiment8,experiment9,experiment10,experiment11,experiment12,experiment13,experiment14,experiment15,experiment16,xie93,r148,dec-models,shelton}
and using molecular dynamics (MD) simulations
\cite{MD,MDb,MD2,MD2b,netz,models,coupling,fabbian,confined-coupling,models-rot,semischematicMCT,MMCT,trans-rot-coupl,masaki}. When
water is cooled, the cooperativity of water molecules increases. Recent
experiments on water show large correlated domains of dipoles at ambient
conditions which have a relaxation time much larger than the autocorrelation
time of individual dipoles \cite{shelton}. MD studies of water models also
show the possibility of formation of large correlated domains of dipoles in
bulk as well as interfacial water \cite{masaki} (where these correlated
patterns of dipoles are pinned to solvated amino acids). These two studies
are the principal motivation for the present investigation of the rotational
cooperativity of water molecules.

A challenging problem is to develop methods of describing molecular
motion in water that are better able to interpret experimental results,
such as incoherent quasielastic neutron scattering, light scattering,
dielectric, and nuclear magnetic resonance experiments
\cite{experiment,xie93}.  Several approximation proposals have been made
for various autocorrelation functions describing both rotational and
translational motion \cite{dec-models,models}. These methods usually
assume the Kohlrausch-Williams-Watts stretched exponential for the long
time relaxation behavior of autocorrelation functions $\phi(t)$, as
predicted by mode coupling theory (MCT),
\begin{equation}
\phi(t)=A \exp\left[-\left(\frac{t}{\tau_{a}}\right)^\beta\right].
\label{stretch}
\end{equation}
The relaxation time $\tau_{a}$, the exponent $\beta$, and the
non-ergodicity factor $A$ are fitting parameters that depend on
temperature $T$ and density $\rho$ \cite{MD,MDb,MD2,MD2b,%
models,coupling,fabbian,confined-coupling,models-rot}.

Our interest here is to study the orientational dynamics of water by simulating SPC/E water. First we calculate the orientational autocorrelation time as the fitting parameter $\tau_{a}$ appearing in Eq.~(\ref{stretch})
\cite{MD,MDb}. Other definitions are possible, e.g., based on other 
fitting functions for the orientational autocorrelation function
decay, such as the biexponential \cite{netz,yeh} or the von Schweidler
law \cite{MMCT}. In all cases, the orientational autocorrelation times are
the result of multi-parameter fitting procedures and roughly correspond to
the characteristic time over which the orientational autocorrelation
function decays by a factor of $e\approx 2.7$.  

To find an upper limit of the orientational autocorrelation time $\tau_{a}$,
we will introduce a new quantity, the dipole randomization time $\tau_{r}$, as the time after which the fluctuations of
the dipoles resemble an uncorrelated random variable \cite{binder}
(Sec.~IV A). We find $\tau_{r}>\tau_{a}$, and that $\tau_{r}$ and
$\tau_{a}$ are linearly related (Sec.~IV B), which is consistent with the MCT
predictions that:
\begin{itemize}

\item[{(i)}] The autocorrelation times of all the autocorrelation functions of any
        fluctuation coupled to density fluctuations diverge at the same
        temperature $T_{\rm MCT}$ with the same power law exponent;

\item[{(ii)}] All the characteristic times of a supercooled liquid are
        proportional to one another. 

\end{itemize}

To characterize the increase of cooperativity and test for the presence of
large correlated domains of dipoles, we also estimate the randomization time
$\tau_{\rm box}$ for the site-dipole field (Sec.~V), a quantity which
measures the relaxation of the net dipole moment of all the molecules inside
a box of edge $L_{\rm box}$. Our calculations show that $\tau_{\rm box}$ when
$L_{\rm box}\approx 3$\AA\ has a power law divergence at $T_{\rm MCT}$, but
with $\tau_{\rm box}<\tau_{r}$. This result shows that the site-dipole field
relaxes faster than the individual dipoles, resolving the apparent
contradiction between Ref.~\cite{masaki} and previous results.  Calculations
of $\tau_{\rm box}$ for larger boxes show that $\tau_{\rm box}$ does not
depend on the box size and hence do not support the experimental observation
of long-lived large domains of correlated dipoles \cite{shelton}.

\section{The SPC/E model}

Our results are based on MD simulations of the extended simple point
charge (SPC/E) model \cite{spce}. The distance between the oxygen atom
and each of the hydrogen atoms is $0.1$~nm, and the HOH angle is the
tetrahedral angle 109.47$^\circ$ \cite{realH2O}.  Each hydrogen atom has
a charge $q_H=0.432e$, where $e$ is the electron charge, and the oxygen
atom has a charge $q_O=-2q_H$. In addition, to model the van der Waals
interaction, pairs of oxygen atoms of different molecules interact with
a Lennard-Jones potential,
\begin{equation}
V_{i,j}(r_{i,j})= 4\epsilon \left[ \left( \frac{\sigma}{r_{i,j}}
  \right)^{12}  -  \left( \frac{\sigma}{r_{i,j}} \right)^{6}  \right],
\end{equation}
where $r_{i,j}$ is the distance between molecules $i$ and $j$,
$\epsilon=0.65$~kJ/mol and $\sigma=0.3166$~nm.

We perform MD simulations for a system of $N=1728$ molecules at density
$\rho = 1.0$~g/cm$^3$, 210~K $\leq T\leq$ 300~K, with periodic boundary
conditions and a simulation time step of 1~fs.  The temperature is
controlled by the Berendsen method of rescaling the velocities
\cite{berendsen}.  The long-range Coulombic interactions
\cite{reactionfield} are treated with the reaction field technique with
a cutoff of $0.79$~nm.  For each state point, we run two independent
simulations to improve statistics.

\section{The orientational autocorrelation function $C_1(t)$}

To estimate the orientational autocorrelation time of water molecules in the
supercooled regime, we average the scalar product of the normalized
dipole vectors $\vec{\mu}_i$ of each water molecule $i$ in the system,
\begin{eqnarray}
C_1(t) &\equiv& 
\left\langle \sum_{i=1}^N \vec{\mu}_i(t) \cdot 
\vec{\mu}_i(0)\right\rangle \nonumber\\
 & = & \frac{1}{N} \sum_{i=1}^N \langle \cos \theta_i(t) \rangle, 
\end{eqnarray}
where $\theta_i(t)$ is the angle between $\vec{\mu}_i(t)$ and
$\vec{\mu}_i(0)$. This function corresponds to the average of the
Legendre polynomial $P_1(\cos\theta_i(t))$ evaluated for each molecule and
can be directly measured by dielectric experiments.

\begin{figure}[p]
\includegraphics[width=11cm]{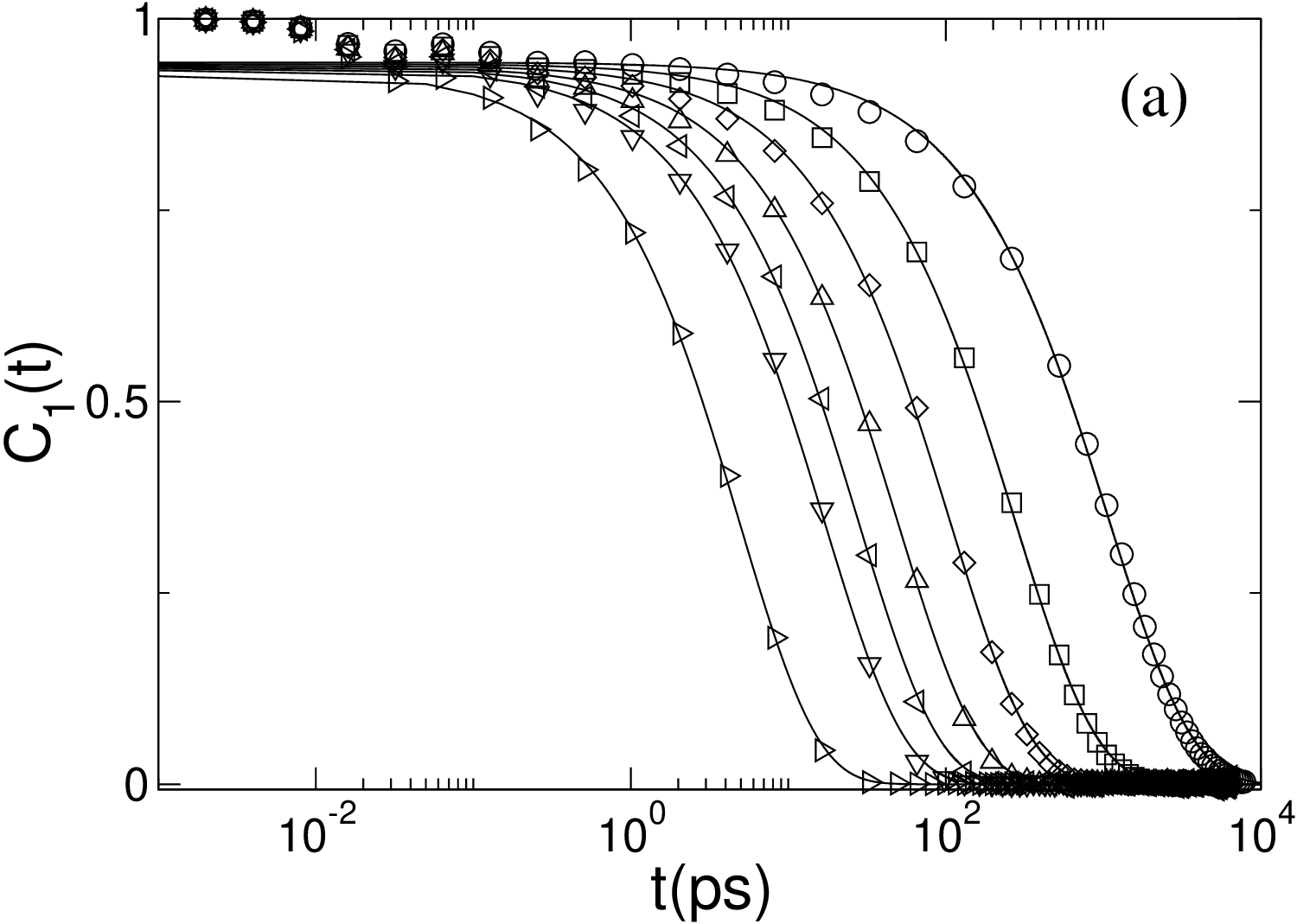}
\includegraphics[width=11cm]{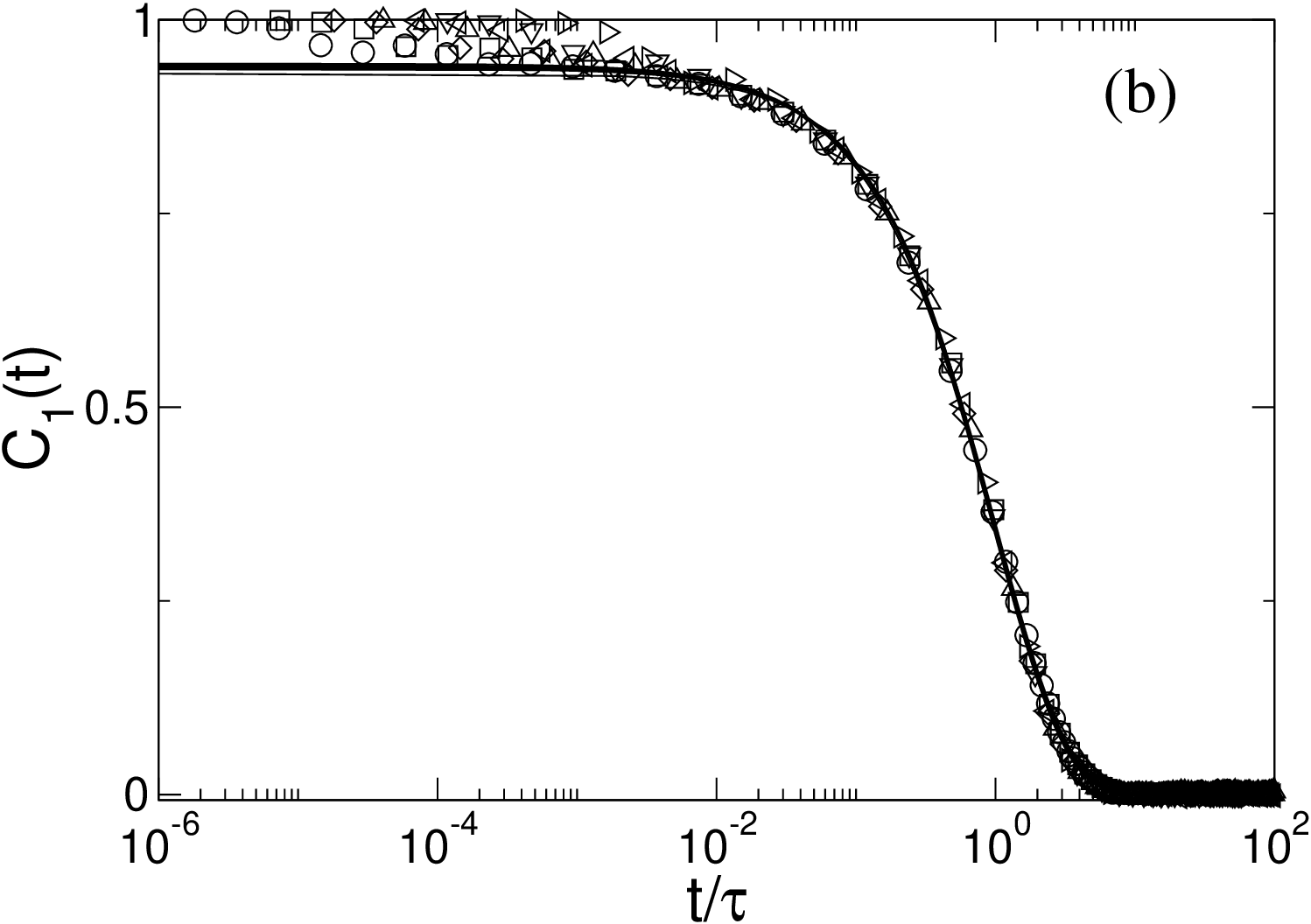}
\caption{(a) The orientational autocorrelation function $C_1$ as a
  function of time $t$ for $T(K)=210$ ($\bigcirc$), 220 ($\Box$), 230
  ($\Diamond$), 240 ($\bigtriangleup$), 250 ($\lhd$), 260
  ($\bigtriangledown$), 300($\rhd$).  Symbols are simulations, lines are
  fits over the range for $t\geq 0.03$~ps to Eq.~(\ref{stretch}) with the fitting
  parameters listed in Table~\ref{fit-par}. (b) Test of
  the time-temperature superposition principle, as predicted by MCT.
  The symbols and the lines for different $T$ fall on a single curve if
  the times are rescaled by $\tau_a(T)$.}
\label{self-dip}
\end{figure}

Figure~\ref{self-dip}(a) plots $C_1(t)$ for 210~K~$\leq T\leq 300$~K,
and displays the two-step decay of typical glass-forming systems. The
long-time regime at low $T$ can be fit well by Eq.~(\ref{stretch}) and
the fitting parameters are shown in Table~\ref{fit-par}.  Both
parameters in Eq.~(\ref{stretch}), $A$ and $\beta$, show weak
dependences on $T$.  The resulting values of these parameters are
consistent with previous simulations of a smaller system of
SPC/E molecules \cite{MDb}.

\begin{figure}[p]
\includegraphics[width=12cm]{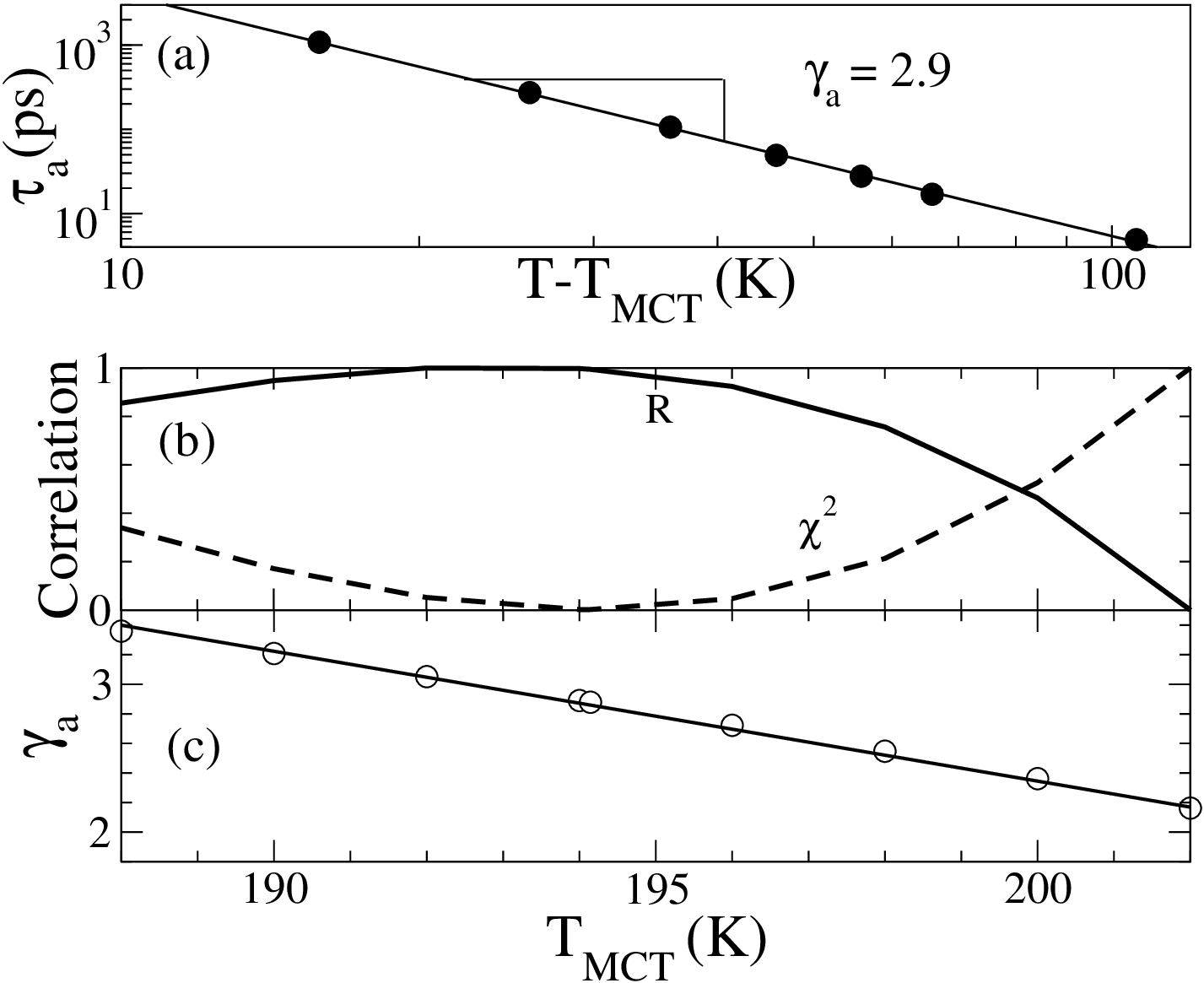}
\caption{(a) Power law behavior of orientational autocorrelation time
$\tau_a$ extracted from $C_1(t)$, as a function of $T-T_{\rm MCT}$, as
predicted by MCT [Eq.~(\ref{power})]. The line is a fit to the MCT power law
with $T_{\rm MCT}=194$~K and exponent $\gamma_a=2.9$. (b) To optimize the fit, we
vary $T_{\rm MCT}$, calculate the autocorrelation coefficient $R$ (solid line)
and the $\chi^2$ (dashed line), and choose as our estimate of $T_{\rm MCT}$
the value corresponding to the maximum or minimum of these quantities, within
a 10\% variation in our range of $T$. $R$ and $\chi^2$ are rescaled
to the maximum and minimum values we found for 188~K $\leq T_{\rm MCT}\leq$
202~K. (c) The MCT exponent $\gamma_a$ corresponding to different choices
of $T_{\rm MCT}$. Note that the exponent $\gamma_a$ decreases almost linearly with
increasing choice of $T_{\rm MCT}$. 
Based on the results in (b), our estimates are $T_{\rm MCT}=(194 \pm 4)$~K
and $\gamma_a=2.9\pm 0.3$.}
\label{powerfit}
\end{figure}

The estimated autocorrelation times $\tau_{a}$ agree (Fig.~\ref{powerfit})
with the power law behavior predicted by the MCT,
\begin{equation}
\tau_{a}\sim (T-T_{\rm MCT})^{-\gamma_a} ~.
\label{power}
\end{equation}
We estimate $T_{\rm MCT}=(194 \pm 4)$~K and $\gamma_a=2.9\pm 0.3$, in agreement
with previous results for similar densities and temperatures \cite{MD2}.

The estimated values of $\tau_{a}$, verify well the the von
Schweidler law (Appendix) and 
the {\it time-temperature superposition principle} predicted by
MCT, i.e. that
the autocorrelation functions in the $\alpha$-relaxation regime at
different temperatures follow the same master curve if the time is
rescaled by the autocorrelation time
(Fig.~\ref{self-dip}b) \cite{fabbian}.

\section{The self-dipole randomization time $\tau_{r}$} 
\subsection{Definition and Methods}

Here we define the randomization time $\tau_{r}$, a new quantity that we
propose to characterize the orientational autocorrelation time.  We consider
the normalized dipole $\vec{\mu}_i$ of molecule $i$ over a time interval
$\Delta t={\cal N}\delta t$,
\begin{equation}
\bar{\mbox{\boldmath $\mu$}}_i \equiv
\frac{1}{{\cal N}} 
\sum_{k=0}^{{\cal N}} \vec{\mu}_i(t_k),
\label{mu-av}
\end{equation}
where \( \bar{\mbox{\boldmath $\mu$}}_i \) is a function of $\delta t$ and
$\Delta t$, $t_k\equiv  
k\delta t$, and $\delta t$ is the time interval between two consecutive
samples of $\vec{\mu}_i$. 

If $\delta t$ is greater than the autocorrelation time of $\vec{\mu}_i$,
then two consecutive samples $\vec{\mu}_i(t)$ and $\vec{\mu}_i(t+\delta t)$
are {\it independent}, hence
$\langle\vec{\mu}_i(t_j) \cdot \vec{\mu}_i(t_k)\rangle=0$ if $j\not=k$,
where $\langle....\rangle$ denotes the average over all the molecules $N$ in
the system. Hence
\begin{equation}
\langle \bar{\mbox{\boldmath $\mu$}}_i^2\rangle \equiv
\langle \bar{\mbox{\boldmath $\mu$}}_i \cdot 
\bar{\mbox{\boldmath $\mu$}}_i
\rangle \equiv
\frac{1}{{\cal N}^2}
\langle \sum_{j,k}^{\cal N}
\vec{{\mu}}_{i}(t_j) \cdot \vec{{\mu}}_{i}(t_k)
\rangle= \frac{1}{\cal N},
\label{mu-sq-av}
\end{equation}
because $\langle (\vec{{\mu}}_{i})^2\rangle=1$ for any $t_k$, and
\begin{equation}
\label{mu-rms}
\mu_{\rm rms} \equiv \sqrt{\langle\bar{\mbox{\boldmath $\mu$}}_i^2\rangle} =
\frac{1}{\sqrt{\cal N}} =
\sqrt{\frac{\delta t}{\Delta t}}~~.
\end{equation}

This is the result of a {\it freely jointed chain\/} of ${\cal N}$ bonds
of the same length, for which the mean square end-to-end distance is
\( {\cal N}^2 \langle\bar{\mbox{\boldmath $\mu$}}_i^2\rangle={\cal N} \)
\cite{feller-flory}. Therefore, if $\delta t$ is larger than the
orientational autocorrelation time for $\vec{\mu}_i$, the
$\mu_{\rm rms}$ decreases as $1/\sqrt{\Delta t}$.

If, instead, $\delta t$ is shorter than the orientational autocorrelation
time, consecutive elements in the sum in Eq.~(\ref{mu-sq-av}) are correlated
$\langle\vec{\mu}_i(t) \cdot \vec{\mu}_i(t+\delta t)\rangle=z$, resulting in
a smaller fluctuation. This can be formally understood by considering the
{\it freely rotating chain\/} model \cite{feller-flory}, where consecutive
bonds in the chain are free to rotate, each around the axis of the previous
bond, at an angle $\theta$, such that $\cos(\theta) = z$. With this
assumption, the resulting mean square end-to-end distance for $n$ bonds of
unit length is
\begin{equation}
\langle r_n^2\rangle= 
n\frac{1+z}{1-z}-2z\frac{1-z^n}{(1-z)^2}.
\label{end-to-end}
\end{equation}
In the case of small $\theta$, we have $z=1-\epsilon+O(\epsilon^2)$,
with $\epsilon=\theta^2/2\ll 1$ and $z^n\simeq
\exp(-n\epsilon)$. Then, from Eq.~(\ref{end-to-end}), we obtain
\begin{equation}
\left\langle \frac{1}{n}\sqrt{r_n^2}\right\rangle= \frac{1}{n\epsilon} 
\left[2\left({n\epsilon}-1-e^{-n\epsilon}\right)\right]^{1/2} ~.
\label{end-to-end-over-n}
\end{equation}

In our problem, the bonds are dipole vectors sampled at time intervals
$\delta t$, and $n = \Delta t/\delta t= {\cal N}$. Therefore
Eq.~(\ref{end-to-end-over-n}) becomes
\begin{equation}
\mu_{\rm rms}\sim
\frac{1}{\cal N\epsilon} 
\left[2\left({\cal N\epsilon}-1-e^{-\cal N\epsilon}\right)\right]^{1/2} ~.
\label{approx-func}
\end{equation}
The right-hand side of this equation behaves as $1/\sqrt{\Delta t}$ for 
${\cal N}\rightarrow \infty$, i.e., the random case behavior is recovered
for large $\Delta t/\delta t$.

Therefore, if we define $\tau_{r}$ as the time at which the
correlation goes to zero as $1/\sqrt{\Delta t}$, it is possible to see
that
\begin{equation}
\mu_{\rm rms}\sim 1/\sqrt{\Delta t} 
\begin{cases}
\mbox{ for any } \Delta t & 
\mbox{if } \delta t\geq \tau_{r}\\
\mbox{ for }\Delta t\gg \tau_{r} & 
\mbox{if } \delta t < \tau_{r}.
\end{cases}
\label{conditions}
\end{equation}
If we consider the fluctuation of any observable, the relation
(\ref{conditions}) defines the randomization time $\tau_{r}$ for that
observable \cite{binder} and $\tau_{r}$ is equal to the smallest $\delta
t$ such that $\mu_{\rm rms}\sim 1/\sqrt{\Delta t}$ for any $\Delta t$.

\subsection{Calculation of $\tau_{r}$}

\begin{figure}[p]
\includegraphics[width=9cm]{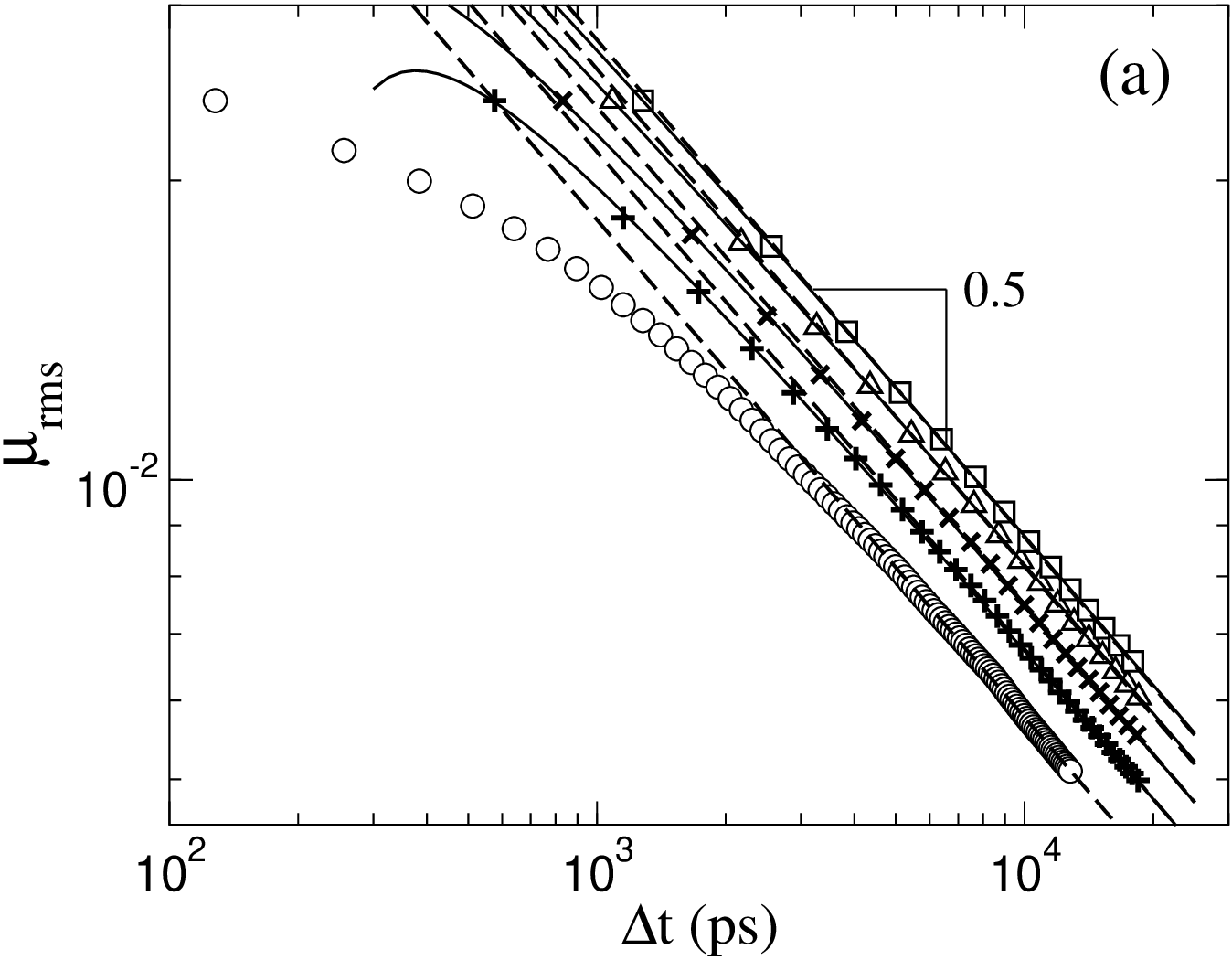}
\includegraphics[width=9cm]{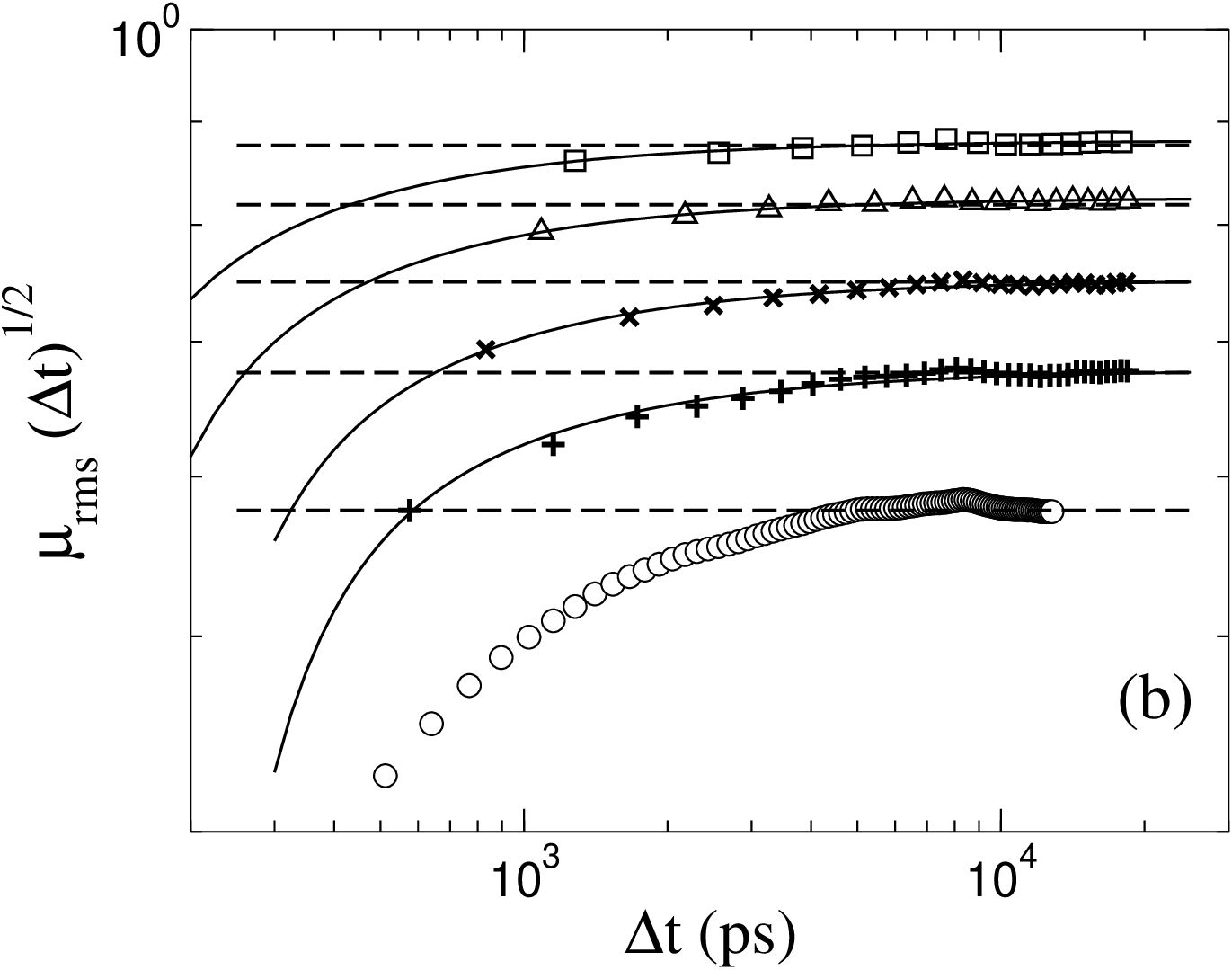}
\caption{(a) The $\mu_{\rm rms}$ of Eq.~(\ref{mu-rms}) for $T=220$~K, plotted
versus $\Delta t$ for a range of different time steps $\delta t=128$~ps
($\bigcirc$), 576~ps ($+$), 832~ps ($\times$), 1088~ps ($\bigtriangleup$),
and 1280~ps ($\Box$). Dashed lines show the predicted asymptotic behavior
$\mu_{\rm rms}\sim 1/\sqrt{\Delta t}$. The fit with Eq.~(\ref{approx-func})
(solid lines) is good when $\delta t\geq 576$~ps, but we are unable to fit
the data for $\delta t=128$~ps, showing that the angle between the dipoles in
Eq.~(\ref{mu-sq-av}) is not independent, as assumed in the freely rotating chain
model. However, Eq.~(\ref{approx-func}) gives a fair description of the
approach to the asymptotic regime.  (b) $\mu_{\rm rms}\sqrt{\Delta t}$ vs $\Delta
t$ approaches a constant asymptotically when $\Delta t\gg\tau_r$.  In
both panels (a) and (b) the errors are roughly the size of the symbols.}
\label{tauhat-self-dip}
\end{figure}

In Fig.~\ref{tauhat-self-dip}, we show $\mu_{\rm rms}$ for $T=220$~K
calculated for different values of $\delta t$. For small $\delta t$ and
small $\Delta t$, $\mu_{\rm rms}$ deviates largely from the asymptotic
law.  However, for increasing $\delta t$, the deviation decreases. For
$\delta t=1280$~ps the asymptotic behavior, within the error of our
calculations, is reached.

\begin{figure}[p]
\includegraphics[width=12cm]{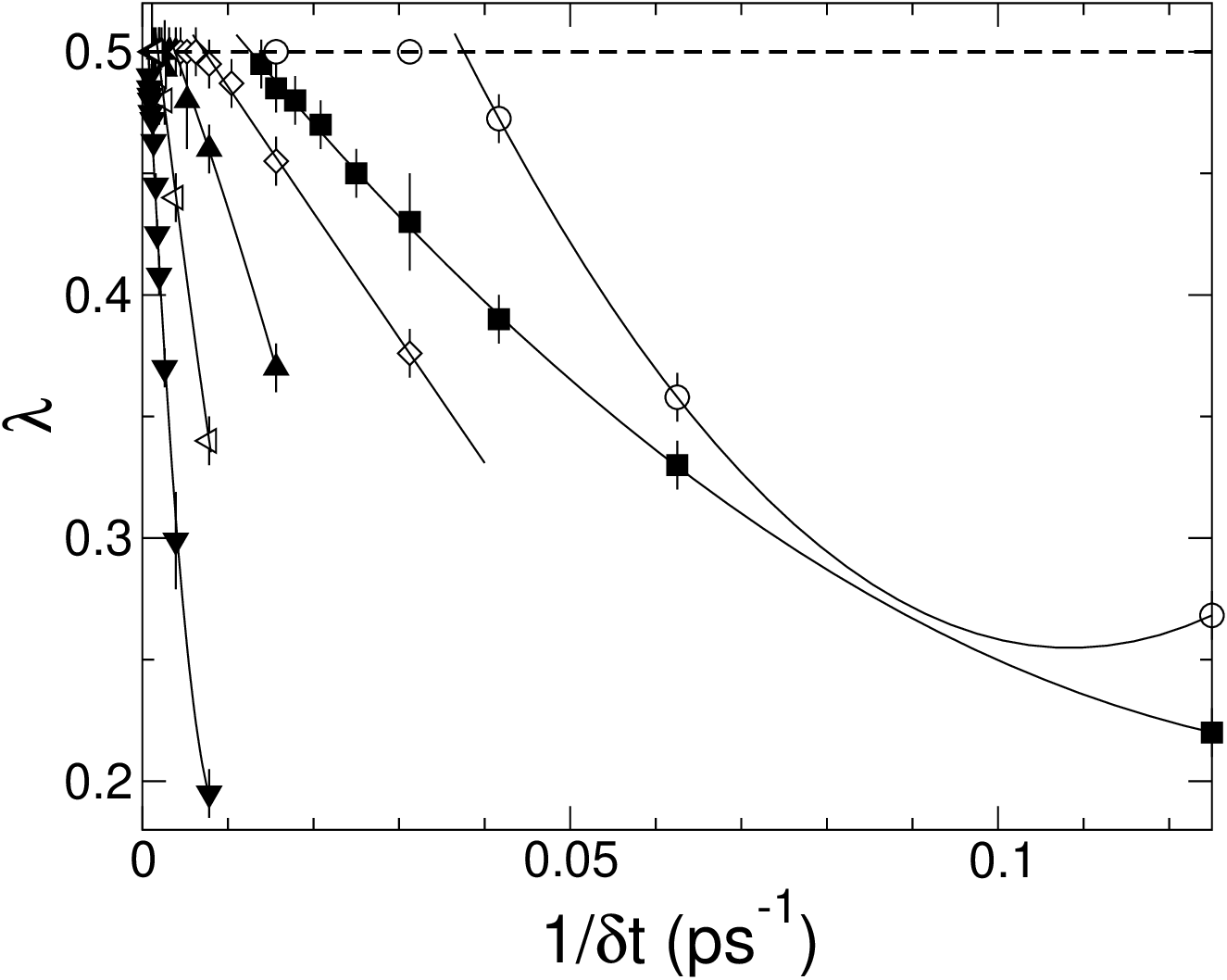}
\caption{The exponent $\lambda$, defined in Eq.~(\ref{lambda}) and calculated
using the first eight points of the curves in Fig.~\ref{tauhat-self-dip},
versus the inverse of time-step $\delta t$, for $T=300$~K ($\bigcirc$), 260~K
($\blacksquare$), 250~K ($\Diamond$), 240~K ($\blacktriangle$), 230~K
($\lhd$), 220~K ($\blacktriangledown$).  Where not shown the errors are
smaller than the symbol size.  The horizontal dashed line corresponds to
$\lambda=0.5$.  By a quadratic fit of the data with $\lambda<0.5$, we find
the self-dipole randomization time $\tau_{r}$, defined as the value of
$\delta t$ where $\lambda=1/2$.}
\label{lambda-fig}
\end{figure}

The evaluation of $\tau_{r}$ from a plot such as in
Fig.~\ref{tauhat-self-dip} could be problematic, since it depends
critically on the data errors. Therefore, to define in a clear way
$\tau_{r}$, we fit the first eight points ($\Delta t = \delta t, 2\delta
t, ....,8\delta t$) using
\begin{equation}
\mu_{\rm rms} \sim
(\Delta t)^\lambda,
\label{lambda}
\end{equation}
where $\lambda=\lambda(\delta t)$.  In this way we study how the
deviation from the asymptotic regime decreases by increasing $\delta
t$. We find that the exponent $\lambda$ increases toward the asymptotic
value $1/2$ for increasing $\delta t$, and $\lambda=1/2$ for any $\delta
t\geq \tau_{r}$ (Fig.~\ref{lambda-fig}). We therefore define $\tau_{r}$
as the extrapolated values of $\delta t$ at which $\lambda=1/2$.  We
find that $\lambda$ approaches 1/2 as $1/\delta t$, to the leading
order, for low temperatures (Fig.~\ref{lambda-fig}).

\begin{figure}[p]
\includegraphics[width=12cm]{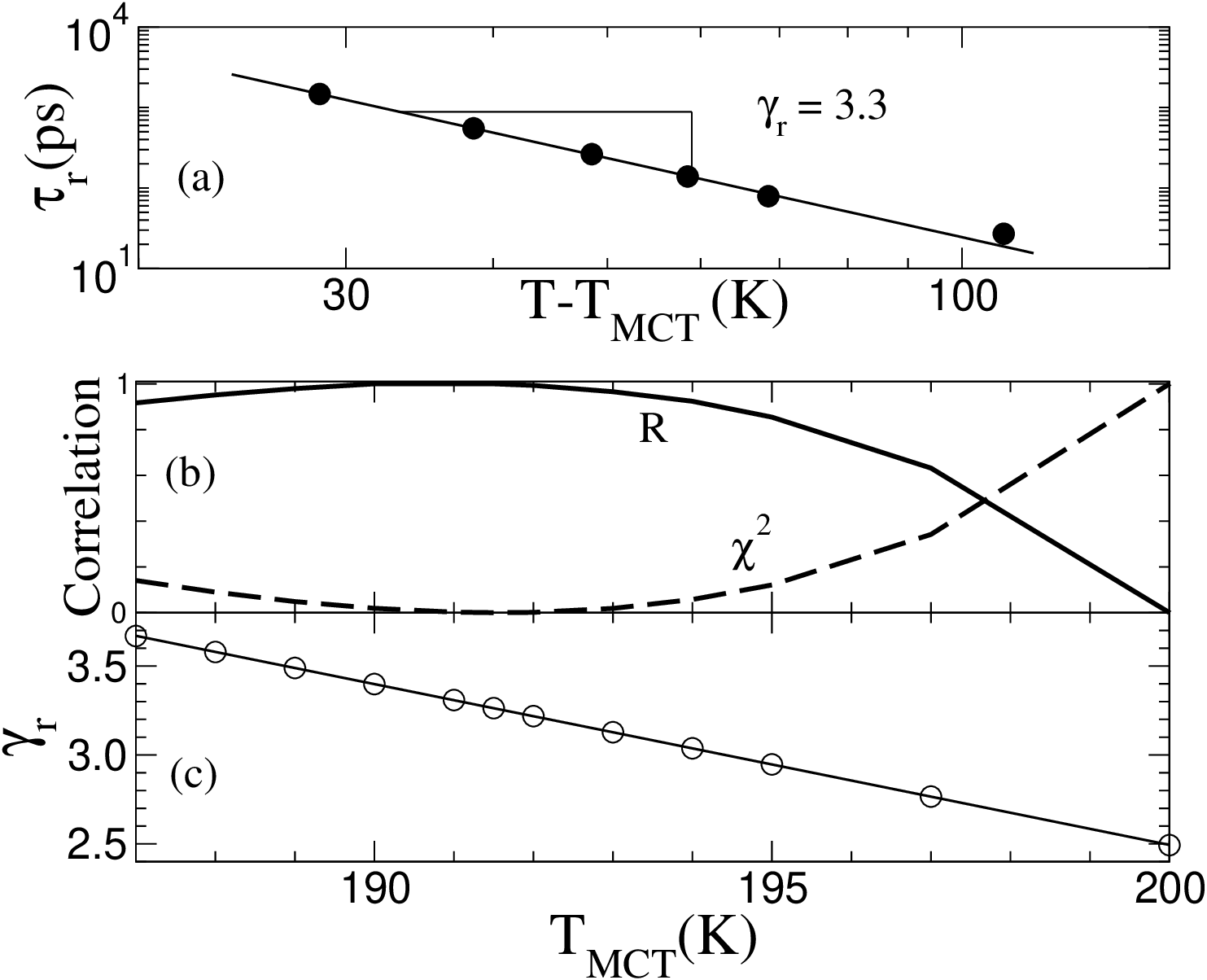}
\caption{Analog of Fig.~2 for the self-dipole randomization time
$\tau_r$. (a) $\tau_{r}$ follows a power law behavior in $T-T_{\rm MCT}$
[Eq.~(\ref{power})], as predicted by MCT. The line is a fit with $T_{\rm
MCT}=191.5$~K and power $\gamma_r=3.3$. (b) As in Fig.~\ref{powerfit},
to optimize the estimate of $T_{\rm MCT}$ we calculate the autocorrelation
coefficient $R$ (solid line) and the $\chi^2$ (dashed line).  In (b),
$R$ and $\chi^2$ are rescaled to the maximum and minimum values we found
for 187~K $\leq T_{\rm MCT}\leq$ 200~K. (c) The fitting parameter
$\gamma_r$ corresponding to different estimates of $T_{\rm MCT}$. The
exponent $\gamma_r$ decreases linearly with increasing estimates of
$T_{\rm MCT}$.
Based on the results in (b), our estimates are $T_{\rm MCT}=(191.5\pm 2.5)$~K
and $\gamma_r=3.3\pm 0.2$.}
\label{tauhat-self}
\end{figure}

The resulting values of $\tau_{r}$ are presented in
Fig.\ref{tauhat-self}a as functions of $T-T_{\rm MCT}$, showing that the
power law behavior Eq.~(\ref{power}) is well satisfied by $\tau_{r}$. In
this case our estimates are $T_{\rm MCT}=(191.5\pm 2.5)$~K and
$\gamma_r=3.3\pm 0.2$, both consistent within the errors with the
estimates based on $\tau_{a}$ (Fig.\ref{powerfit}).  Therefore, the
prediction (i) of MCT is verified.

By plotting $\tau_{r}$ against $\tau_{a}$, we verify the MCT prediction
(ii). We find (Fig.~\ref{tau-tauhat}) that $\tau_{r}$ and $\tau_{a}$
are linearly related and that $\tau_{r}$ is approximately five times
larger than $\tau_{a}$.

The large value of $\tau_{r}$ with respect to $\tau_{a}$ is consistent
with the fact that the latter measures the decay of the self-dipole
correlation to a finite value, while the first measures the time needed
for the self-dipole autocorrelation to decay to zero.  This result is also
reminiscent of the recent MD analysis in bulk water for the {\it
site-dipole field}, a measure of the average orientation of the
molecules passing through each spatial position, recently introduced in
Ref.~ \cite{masaki}.  Higo et al. \cite{masaki} find coherent patterns
for the site-dipole field, at ambient pressure and $T=298$~K and
$T=300$~K, that persist for more than 100~ps, a time much larger than
the single molecule orientational relaxation time $\tau_{a}$ of
approximately 5~ps (Table I).  It is, therefore, interesting to
calculate the randomization time $\tau_{\rm box}$ and to find its
relation with the autocorrelation time $\tau_{a}$ for $T\to T_{\rm
MCT}$.

\section{The site-dipole field}

To check if there are large correlated domains of dipoles in water which
have large relaxation times compared to the individual dipole
correlation time, we next study site-dipole field introduced by Higo
et. al. \cite{masaki}.  We define the instantaneous coarse-grained
site-dipole field
\begin{equation}
\vec{d}^{v}_{i}\equiv \vec{d}(\vec{r_i},t)\equiv
\frac{1}{n_i(t)}\sum_{\rm box}\vec{\mu}_i
\label{cgsd}
\end{equation}
as the average of dipoles $\vec{\mu}_i$ of all the molecules $n_i(t)$ at time
$t$ belonging to box $i$ of edge $L_{\rm box}$, volume 
${v}=L^3_{\rm box}$ and
centered at $\vec{r_i}$. If $n_i(t)=0$, then $\vec{d}^{v}_i=0$ by
definition \cite{note2,mathias}.  We chose vectors $\vec{r_i}$ 
in such a way that the corresponding boxes do not overlap \cite{note1}. The
time average $\bar{d}^{v}_i$ over an interval
$\Delta t$, is defined analogously to Eq.~(\ref{mu-av}). The rms average $d^{v}_{{\rm rms}}$, is defined in analogy to Eq.~(\ref{mu-sq-av}) and (\ref{mu-rms}),
but instead of summation over all molecules we perform a summation over all
boxes.

\begin{figure}[p]
\includegraphics[width=12cm]{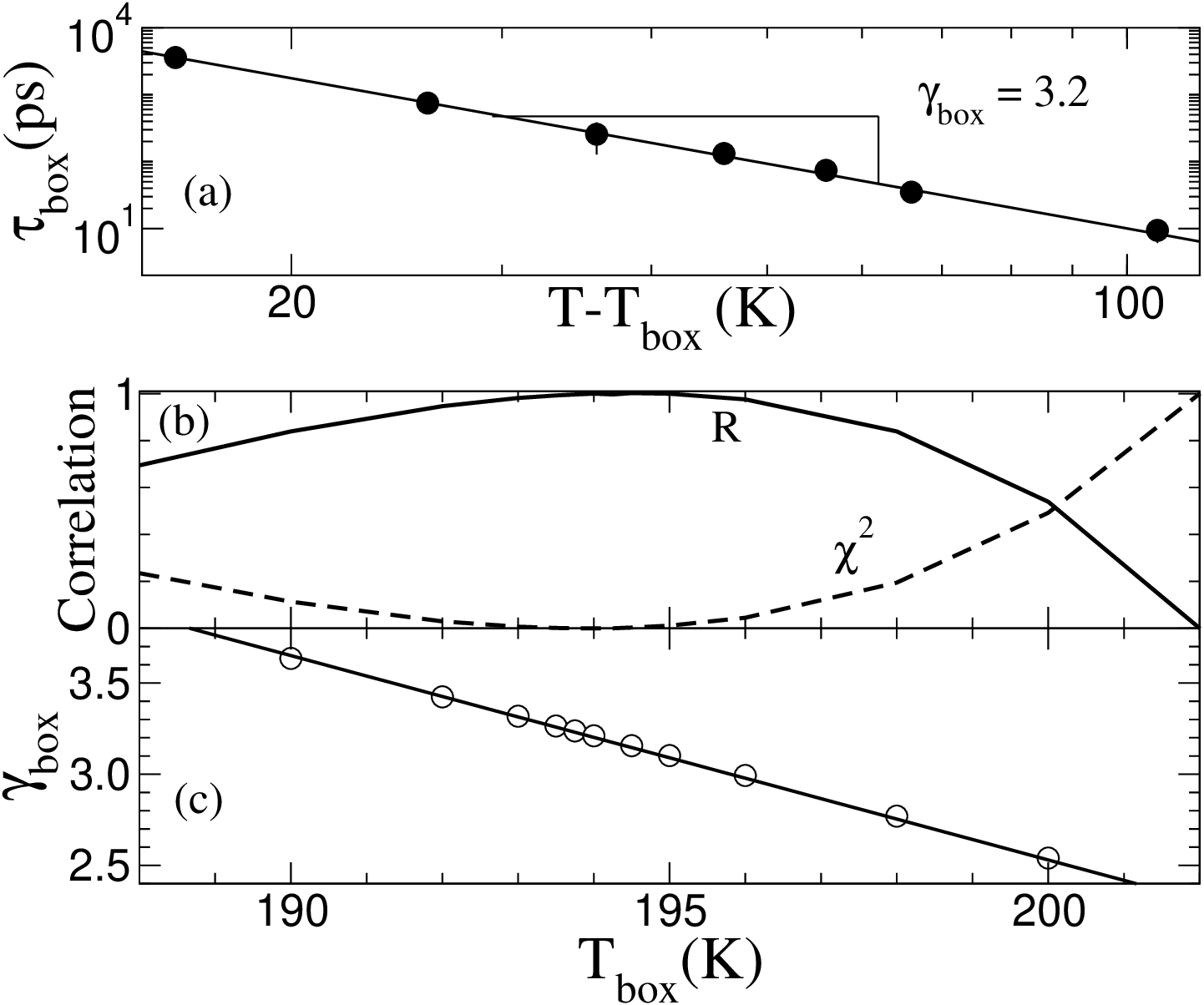}
\caption{Analog of Fig.~2 and Fig.~6 for the site-dipole randomization time
$\tau_{\rm box}$. (a) We find a power law behavior in $T-T_{\rm box}$, calculated
for $L_{\rm box}=3$\AA. The line is a fit with $T_{\rm MCT}=194$~K and
exponent $\gamma_{\rm box}=3.2$. (b) Optimization analysis for $T_{\rm box}$:
correlation coefficient $R$ (solid line) and $\chi^2$ (dashed line), both
rescaled to the maximum and minimum values found for $188$~K$\leq T_{\rm
box}\leq 202$~K. (c) The exponent $\gamma_{\rm box}$ corresponding to
different choices of $T_{\rm box}$, decreases linearly with increasing choice of
$T_{\rm box}$. 
We estimate $T_{\rm box}=(194\pm 2)$~K and
$\gamma_{\rm box}=3.2\pm 0.2$.}
\label{tauhatR3}
\end{figure}

Since the argument presented for $\mu_{\rm rms}$ is also valid for $d^{v}_{{\rm
rms}}$, the relation (\ref{conditions}) also holds for $d^{v}_{{\rm rms}}
$ and allows us to estimate the randomization time $\tau_{\rm box}$ for $d^{v}_{{\rm rms}}$. We find that $\tau_{\rm box}$, calculated for $L_{\rm
box}=3.33$\AA~, diverges at $T_{\rm box}=(194\pm 2)$~K with a power
law with exponent $\gamma_{box}=3.2\pm 0.2$, consistent with our estimates of
$\gamma_a$ and $T_{\rm MCT}$, respectively (Fig.~\ref{tauhatR3}).

\begin{figure}[p]
\includegraphics[width=12cm]{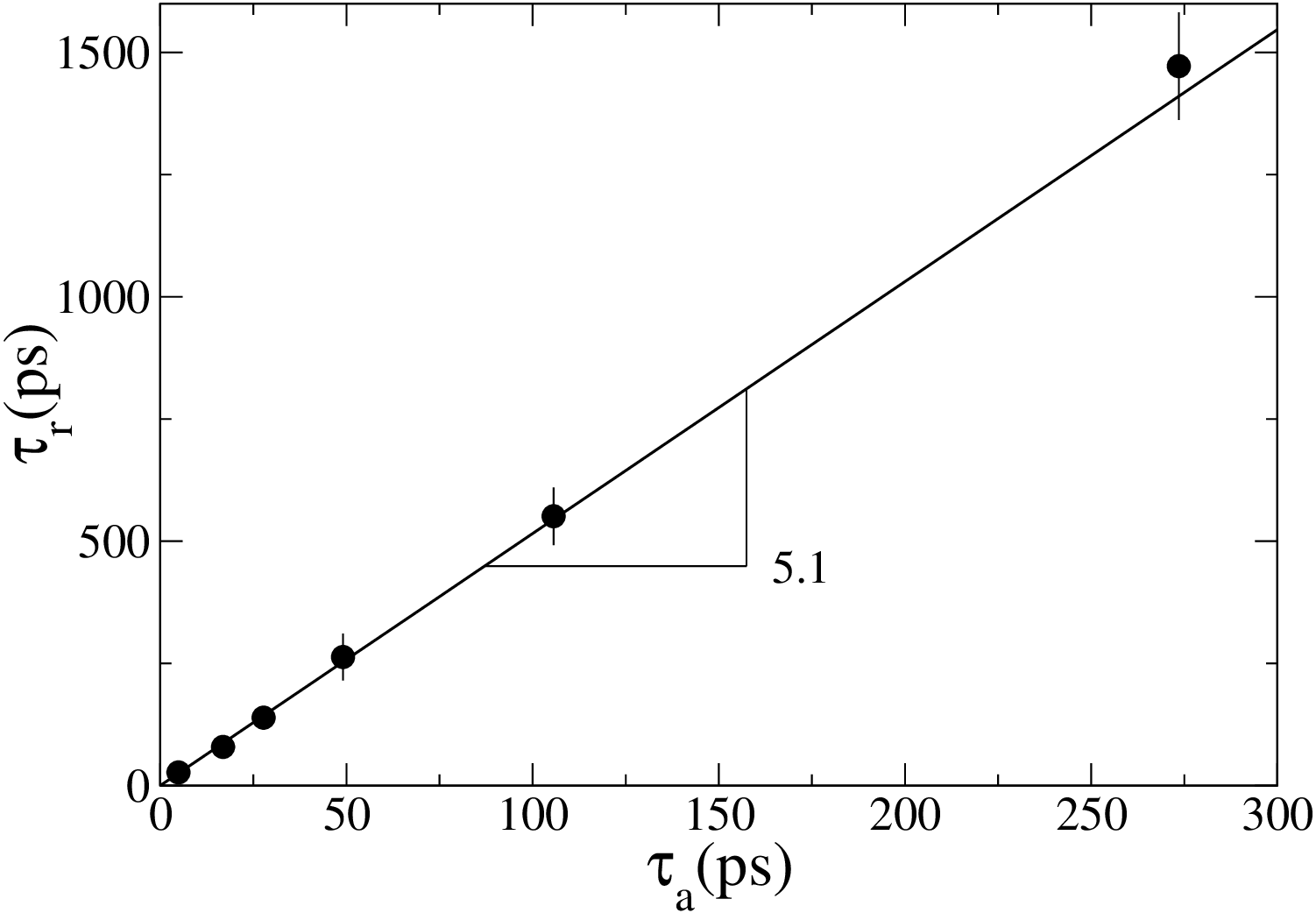}
\caption{Parametric plot of the times $\tau_{r}(T)$ and $\tau_a(T)$, within
  the range $220$~K~$\leq T\leq 300$~K, with the lowest time corresponding to
  the highest $T$. The line reflects the linear one-parameter fit,
  $\tau_{r}=(5.1\pm 0.2) \tau_a$.}
\label{tau-tauhat}
\end{figure}

\begin{figure}[p]
\includegraphics[width=12cm]{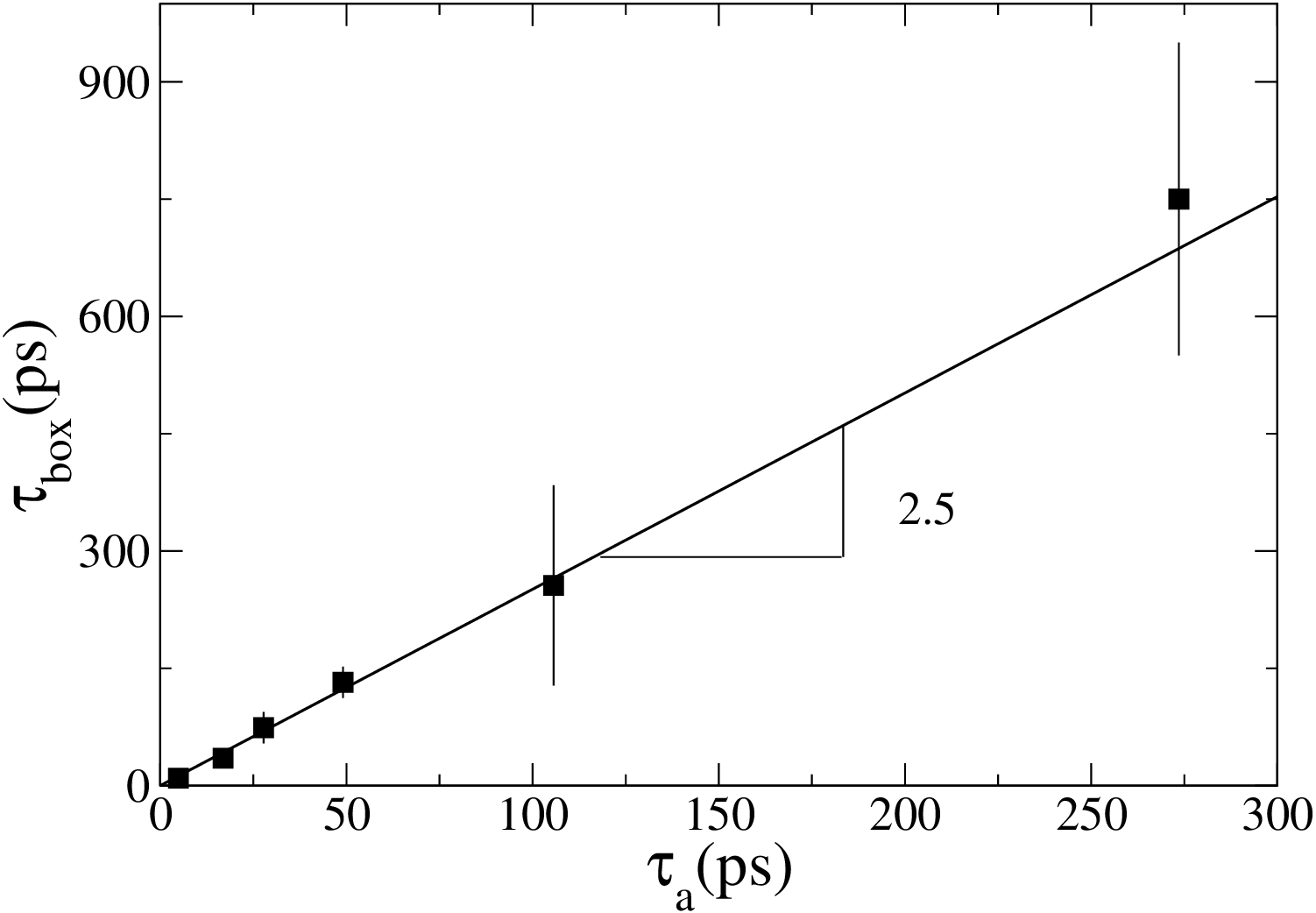}
\caption{Analog of Fig.~7: A parametric plot of the site-dipole
  randomization time $\tau_{\rm box}(T)$ and the orientational autocorrelation time
  $\tau_a(T)$ over the range $220$~K~$\leq T\leq 300$~K, with the lowest
  time corresponding to the highest $T$. The line reflects the linear
  one-parameter fit $\tau_{\rm box}=(2.5\pm 0.2) \tau_a$.}
\label{tau-tauhatR3}
\end{figure}

If we compare $\tau_{\rm box}$ with $\tau_{r}$ (Fig.~\ref{tau-tauhatR3}), we
again find a linear relation, as in Fig.~\ref{tau-tauhat} for $\tau_a$,
consistent with the MCT statement (ii).  The proportionality factor is
approximately 2.5 \cite{note3}, smaller than the factor $\approx 5$ found for
$\tau_{r}$ in Fig.~\ref{tau-tauhat}. Therefore, we conclude that in bulk
water the coarse-grained site-dipole randomization time $\tau_{\rm box}$ is
larger than the self-dipole autocorrelation time $\tau_a$, but smaller than
$\tau_{r}$. Thus we do not find a significant increase in the box dipole
autocorrelation time compared to the autocorrelation time $\tau_a$. These
results do not support the results of Refs. \cite{shelton,masaki}.

\begin{figure}[p]
\includegraphics[width=10.5cm]{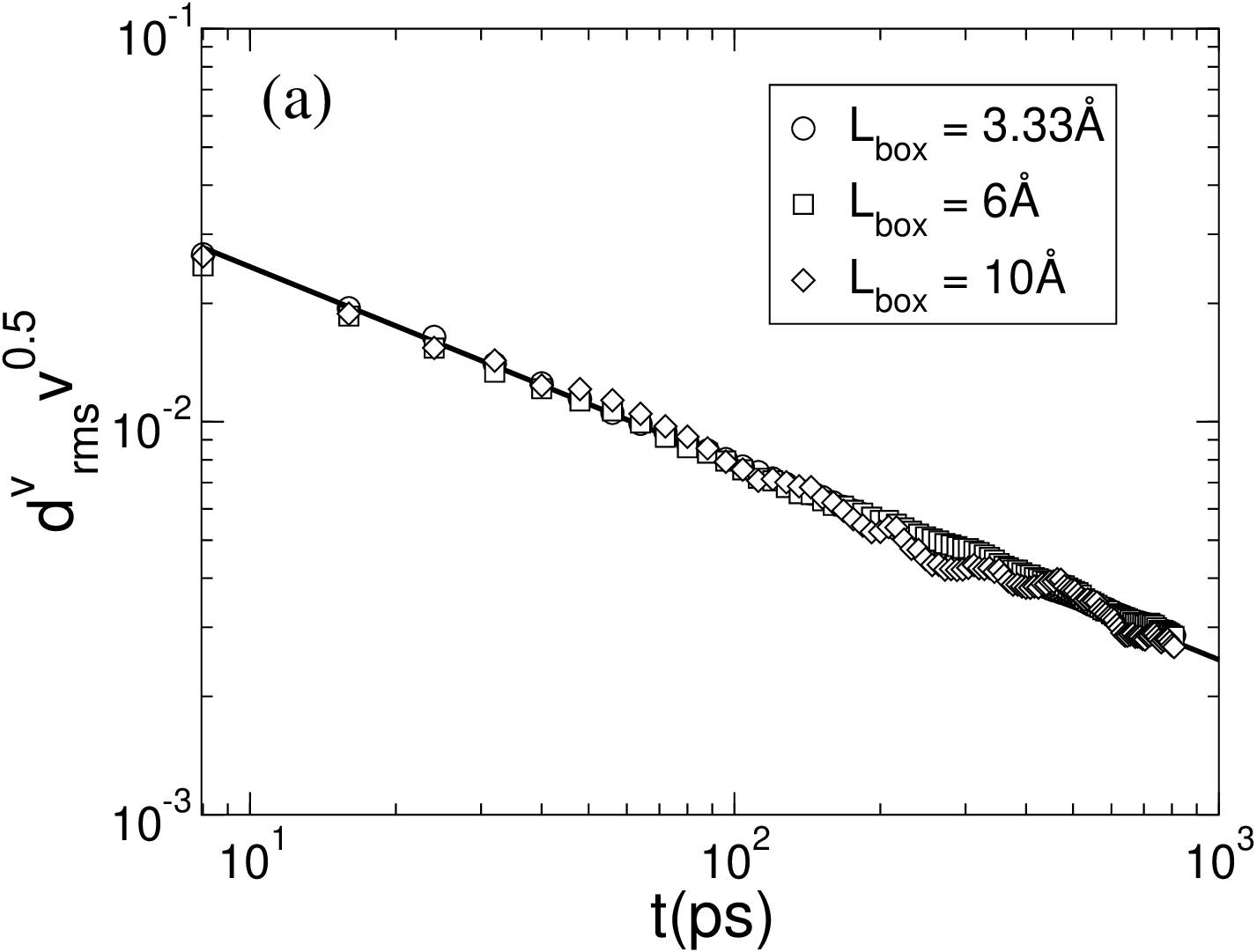}
\includegraphics[width=10cm]{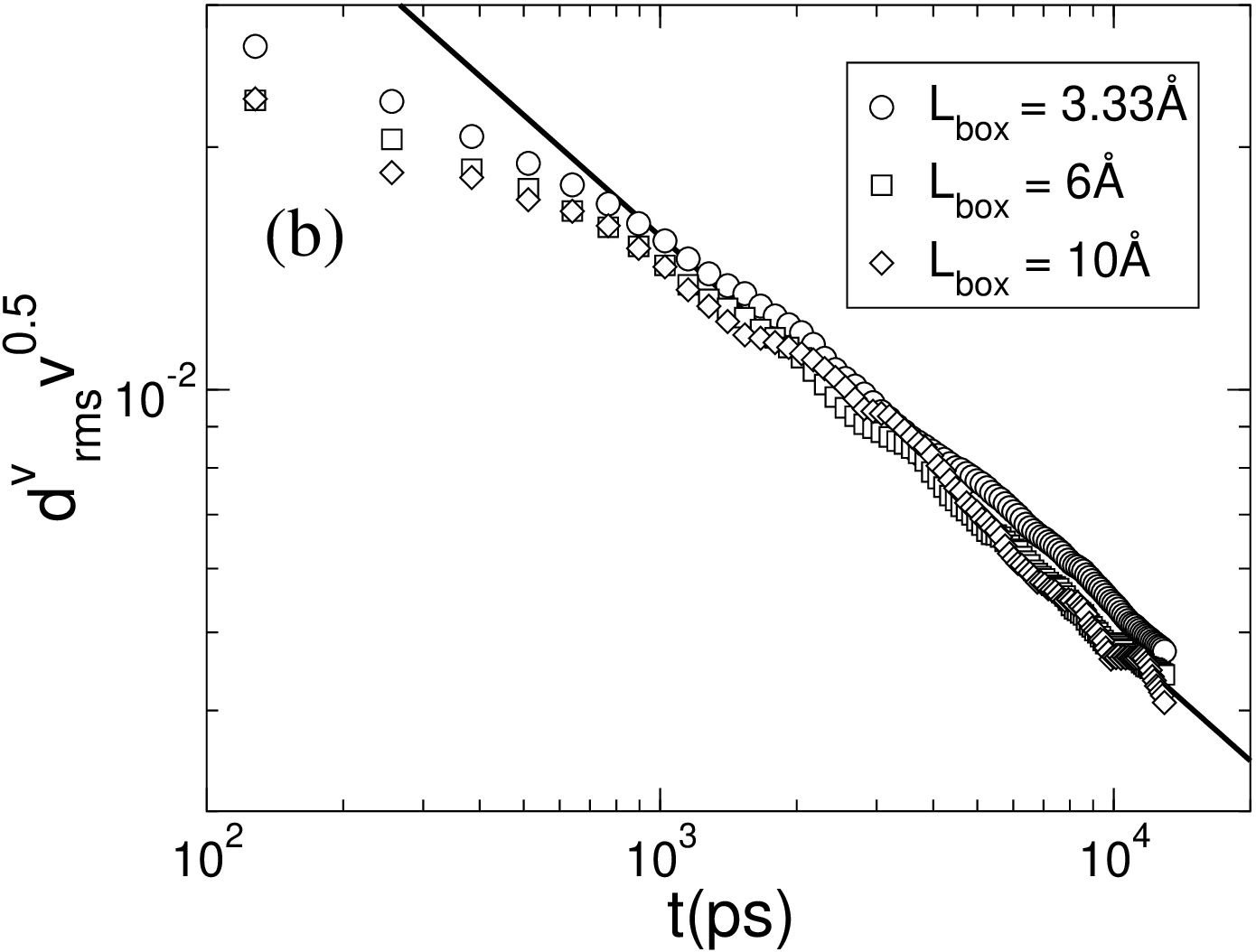}
\caption{Size dependence of site-dipole autocorrelation function $d^{v}_{\rm
rms}$ as a function of $t$ for $L_{\rm box}=3.33$~\AA~ ($\bigcirc$),
6~\AA~ ($\Box$), 10~\AA~ ($\Diamond$) and for two different temperatures
(a) $T=300$~K and (b) $T=220$~K. In (a) the line is a fit of data for
$L_{\rm box}=3.33$~\AA~ with $d^{v}_{\rm rms} =a/\sqrt{t}$, with
$a=0.08\pm 0.01$. In (b) the same fit is for the data at $L_{\rm
box}=6$~\AA~ and $t>10^2$, with $a=0.49\pm 0.01$.  For each $T$ all the
values of $d^{v}_{\rm rms}\sqrt{v}$ overlap, suggesting that the
orientational autocorrelation is short-range.}

\label{new-domain}
\end{figure}

To test the existence of cooperative domains in the SPC/E model, we perform
coarse-graining of the dipole field for boxes of sizes
$3.33$~\AA~$\leq L_{\rm box}\leq 10$~\AA.  If the dipoles of molecules
in the box are independent 
random variables, $d^{v}_{\rm rms}$ must be inversely proportional to
$\sqrt{<n_i>} \propto \sqrt{v}$, since the average number of molecules
in the box is proportional to its volume. The dependence of $d^{v}_{\rm rms}
\sqrt{v}$ on time $t$ must be the same for the boxes of different
volumes $v$. We show in Fig.~\ref{new-domain} the behavior of $d^{v}_{\rm rms} \sqrt{v}$ for $T=220$~K and $T=300$~K. The collapse of all the
curves confirms the hypothesis of very weak autocorrelations among neighboring
dipoles. Only for $T=220$~K we observe a weak size dependence of $d^{v}_{\rm rms} \sqrt{v}$ for the smallest size, suggesting that at this $T$ the
correlation length is between $3.33$ and $6$~\AA, comparable to the
dipole-dipole correlation length at ambient $T$ \cite{mathias}.  Thus our
simulations support the existence of only short range orientational
autocorrelation in SPC/E water even at low $T$.

\section{Discussion}

Considerable numerical evidence shows that MCT predictions apply to
orientational dynamics of water, despite the fact that MCT has been
developed for particles interacting through spherically symmetric potentials
\cite{MCT}. However, recent extensions of MCT to liquids of linear
molecules \cite{Schilling97,Kammerer97}, and single solute molecules in a
simple solvent liquid \cite{orientational}, confirm the main MCT
predictions about the orientational autocorrelation functions \cite{MMCT}.

Our study of supercooled water confirms the validity of MCT predictions for
the orientational autocorrelation time $\tau_a$, estimated through a stretched
exponential of the dipole autocorrelation function, for the temperature range
$210$~K$\leq T\leq 300$~K at density $\rho=1$~g/cm$^3$. Our results agree
with the time-temperature superposition principle and the power law
Eq.~(\ref{power}), with $T_{\rm MCT}= (194\pm 4)$~K and $\gamma_a=2.9\pm 0.3$.

By evaluating the randomization time $\tau_{r}$, defined as the time needed
to randomize the molecular dipoles, we verify the MCT prediction that all the
characteristic times of quantities coupled to density fluctuations of a
supercooled liquid are proportional to each other and follow the same power
law Eq.~(\ref{power}). We find $\tau_{r}\approx 5~\tau_{a}$, with $T_{\rm
MCT}= (191.5 \pm 2.5)$~K and $\gamma_r=3.3\pm 0.2$, consistent with the
estimates based on the calculation of $\tau_{a}$.

We also calculate the randomization time $\tau_{\rm box}$ for the box dipole
field, a quantity introduced in Ref.~\cite{masaki} to measure the local
orientational memory of molecules passing through a given spatial position.
Our results  for $L_{\rm box}=3.33$~\AA show that $\tau_{\rm box}$ diverges at $T_{\rm box}=T_{\rm MCT}$,
following a power law with exponent $\gamma_{\rm box}=\gamma_a$, and that
$\tau_{\rm box}\approx \tau_{r}/2$. As a
consequence, the local memory is lost faster than the self-dipole orientational memory.

Our results also show the existence of domains of correlated dipoles of short
spatial range, with a correlation length comparable to the dipole-dipole
correlation length at ambient $T$ \cite{mathias}. Whether this conclusion
is specific to the SPC/E model with reaction field is an open question, and
requires further investigation using other models of water, e.g. polarizable models.

\subsection*{Acknowledgments}
We thank N. Giovambattista for his helpful collaboration on initial
phase of this work. We thank the NSF Chemistry Program for support.
GF thanks the Spanish Ministerio de Educaci\'on y Ciencia (Programa
Ram\'on y Cajal and Grant No. FIS2004-03454), and M. Sasai for his
hospitality during a visit to Nagoya University.

\appendix
\section{The von Schweidler law}

\begin{figure}[p]
\includegraphics[width=12cm]{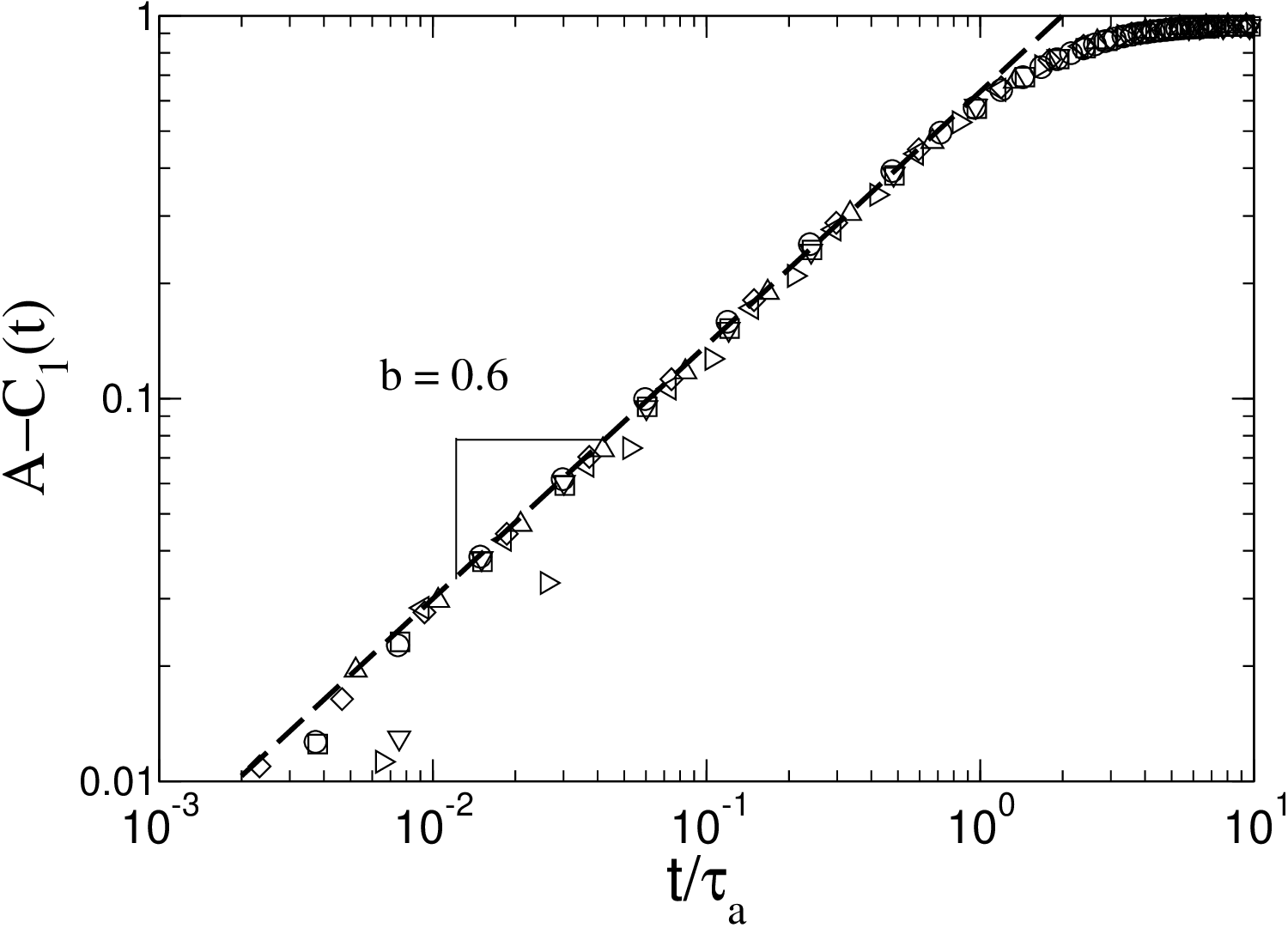}
\caption{Test of the von Schweidler law, Eq.~(\ref{vonSchweidler}). It is
well verified roughly over two decades, for $T\leq 250$~K. Data for higher
$T$ depart from this law at short times. Symbols are as in
Fig.~\ref{self-dip}. Dashed line is the fit to Eq.~(\ref{vonSchweidler}) of
the data for $T=230$~K over the fitting range $0.01\leq A-C_1\leq 0.50$ with
the result $b=0.6\pm 0.1$.}
\label{vonSchweidlerfit}
\end{figure}

The MCT predicts that the autocorrelation function departs from the
plateau $A$ as a power law with exponent $b$, known as the von
Schweidler law,
\begin{equation}
C_1(t)-A\sim -(t/\tau_{a})^b ~,
\label{vonSchweidler}
\end{equation}
where the von Schweidler exponent $b$ does not depend on $T$.  We verify
that at lower temperatures Eq.~(\ref{vonSchweidler}) holds for roughly
two decades in time (Fig.~\ref{vonSchweidlerfit}) and we find a clear
deviation only for $T \ge 260$~K at short times, possibly due to the
fact that for $T\geq 260$~K it is more difficult to estimate the plateau
$A$. The estimated value of $b$ is $0.6 \pm 0.1$, consistent with
previous results \cite{fabbian} and with the MCT prediction that
$\gamma_a$, $a$, and $b$ are related by the equation
\begin{equation}
\gamma_a=\frac{1}{2a}+\frac{1}{2b}.
\label{eq1}
\end{equation}
Here $a$ is the exponent of the power law that describes the short-time
approach to the plateau $C_1-A\sim t^{-a}$, and $a$ is related to $b$ by
the transcendental equation
\begin{equation}
\frac{[\Gamma(1-a)]^2}{\Gamma(1-2a)}=\frac{[\Gamma(1+b)]^2}{\Gamma(1+2b)},
\label{eq2}
\end{equation}
where $\Gamma(x)$ is the Euler gamma function. Our estimates of $b$ and
$\gamma_a$ are consistent with both Eqs.~(\ref{eq1}) and (\ref{eq2})
with $a=0.25 \pm 0.05$.

The values of the exponents $a$, $b$ and $\gamma_a$ are not universal, but
depend on density.  However, the rescaling of the autocorrelation functions
for different $T$ on the same master curve, shows that the orientational
correlation function depends on $T$ and $\rho$ only through the
dependence on $\tau_{a}$, as predicted by the MCT.

\newpage

\begin{table}[hp]
\caption{Parameters of the fit of $C_1(t)$ in Fig.~\ref{self-dip} with
Eq.~(\ref{stretch}).  The error on each parameter is $\pm 10\%$.}
\begin{ruledtabular}
\begin{tabular}{llll}
$T($K$)$ & $A$ & $\tau_a($ps$)$ & $\beta$ \\
\hline
300 & 0.93  & $4.9\times 10^0$  & 0.88 \\
260 & 0.94  & $1.7\times 10^1$  & 0.85 \\
250 & 0.94  & $2.8\times 10^1$  & 0.85 \\
240 & 0.94  & $4.9\times 10^1$  & 0.84 \\
230 & 0.94  & $1.1\times 10^2$  & 0.84 \\
220 & 0.94  & $2.7\times 10^2$  & 0.83 \\
210 & 0.94  & $1.1\times 10^3$  & 0.82 \\
\end{tabular}
\end{ruledtabular}
\label{fit-par}
\end{table}

\newpage

\newpage

\newpage

\newpage

\newpage

\newpage

\newpage

\newpage

\newpage 

\newpage

\begin{references}

\bibitem{review} See, e.g., M.-C. Bellissent-Funel, ed., {\it Hydration
Processes in Biology: Theoretical and Experimental Approaches\/} (IOS Press,
Amsterdam, 1999); P. G. Debenedetti, J. Phys. Cond. Matt. {\bf 15}, R1669
(2003); P. G. Debenedetti and H. E. Stanley, Physics Today {\bf 56} [issue
6], 40--46 (2003); O. Mishima and H. E. Stanley, Nature {\bf 396}, 329--335
(1998).

\bibitem{experiment}
G. Sposito, J. Chem. Phys. {\bf 74}, 6943 (1981).

\bibitem{experiment2}
S. H. Chen, J. Teixeira, and R. Nicklow, Phys. Rev. A {\bf 26}, 3477
(1982).

\bibitem{experiment3}
 F. X. Prielmeir, E. W. Lang, R. J. Speedy, and H. D. L\"{u}deman,
 Phys. Rev. Lett. {\bf 59}, 1128 (1987).

\bibitem{experiment4}
C. A. Angell, Nature {\bf 331}, 206 (1988).

\bibitem{experiment5}
V. Mazzacurati, A. Nucara, M. A. Ricci, G. Ruocco, and G. Signorelli,
J. Chem. Phys. {\bf 93}, 7767 (1990).

\bibitem{experiment6}
F. Sobron, F. Puebla, F. Rull, and O. F. Nielsen,
Chem. Phys. Lett. {\bf 185}, 393 (1991).

\bibitem{experiment7}
K. Mizoguchi, Y. Hori, and Y. Tominaga, J. Chem. Phys.
{\bf 97}, 1961 (1992).
%

\bibitem{experiment8} S.-H. Chen, in {\it Hydrogen-Bonded Liquids},
Vol. 329 of NATO Advanced Studies Institute Series, edited by J.
C. Dore and J. Teixeira (Kluwer Academic, Dordrecht, 1991),
pp.~289--332;
S. Dellerue and M.-C. Bellissent-Funel,
Chem. Phys. {\bf 258}, 315 
(2000);
A. R. Bizzarri and S. Cannistraro,
J. Phys. Chem. B {\bf 106}, 6617 
(2002).
 
\bibitem{experiment9}
S.-H. Chen, P. Gallo, and M.-C. Bellissent Funel, Can. J. Phys. {\bf
  73}, 703 (1995). 


\bibitem{experiment10}
M.-C. Bellissent-Funel, S.-H. Chen, and J.-M. Zanotti, Phys. Rev. E
{\bf 51}, 4558 (1995).

\bibitem{experiment11}
J.-M. Zanotti, M.-C. Bellissent-Funel, and S.-H. Chen, Phys. Rev. E
{\bf 59}, 3084 (1999).

\bibitem{experiment12}
S. Takahara, M. Nakano, S. Kittaka, Y. Kuroda, T. Mori, H. Hamano, and
T. Yamaguchi, J. Phys. Chem. {\bf 103}, 5814 (1999). 


\bibitem{experiment13}
J. T. Cabral, A. Luzar, J. Teixeira, and M.-C. Bellissent-Funel,
J. Chem. Phys. {\bf 113}, 8736 
(2000).


\bibitem{experiment14}
V. Crupi, D. Majolino, P. Migliardo, and V. Venuti,
J. Phys. Chem. A {\bf 104}, 11000 
(2000).


\bibitem{experiment15}
V. Crupi, D. Majolino, P. Migliardo, and V. Venuti,
J. Chem. Phys. B {\bf 106}, 10884 
(2002).


\bibitem{experiment16}
S. Magazu and G. Maisano, J. Mol. Liq. {\bf 93}, 7 (2001).

\bibitem{xie93} Y. Xie, K. F. Ludwig, Jr., G. Morales, D. E. Hare, and
C. M. Sorensen, Phys. Rev.Lett. {\bf 71}, 2050 (1993).

\bibitem{r148} L. Bosio, J. Teixeira, and H. E. Stanley,
  Phys. Rev. Lett. {\bf 46}, 597 (1981). 



\bibitem{dec-models}
J. Teixeira, M.-C. Bellissent-Funel, S.-H. Chen, and A. J. Dianoux,
Phys. Rev. A {\bf 31}, 1913 (1985).

\bibitem{shelton}
D. P. Shelton, Phys. Rev. B, {\bf 72}, 020201 (2005).

\bibitem{MD}
 P. Gallo, F. Sciortino, P. Tartaglia, and S.-H. Chen,
 Phys. Rev. Lett. {\bf 76}, 2730 (1996).

\bibitem{MDb}
 F. Sciortino, P. Gallo, P. Tartaglia, and S.-H. Chen, Phys. Rev. E
 {\bf 54}, 6331 (1996).

\bibitem{MD2}
F. W. Starr, S. Harrington, F. Sciortino, H. E. Stanley, 
Phys. Rev. Lett. {\bf 82}, 3629 
(1999);
F. W. Starr, F. Sciortino, and H. E. Stanley, 
Phys. Rev. E {\bf 60}, 6757 
(1999).

\bibitem{MD2b}
D. Paschek and A. Geiger,
J. Phys. Chem. B {\bf 103}, 4139 
(1999);
P. Gallo, M. Rovere, and E. Spohr, Phys. Rev. Lett. {\bf 85}, 4317 
(2000).

\bibitem{netz}
P. A. Netz, F. W. Starr, H. E. Stanley, and M. C. Barbosa,
J. Chem. Phys. {\bf 115}, 344 
(2001);
P. A. Netz, F. W. Starr, M. C. Barbosa, and H. E. Stanley,
J. Mol. Liq. {\bf 101}, 159 
(2002).

\bibitem{models}
 S.-H. Chen, C. Liao, F. Sciortino, P. Gallo, and P. Tartaglia,
Phys. Rev. E {\bf 59}, 6708 (1999);
C. Y. Liao, F. Sciortino, and S. H. Chen, Phys. Rev. E {\bf 60}, 6776 
(1999).

\bibitem{coupling}
D. Di Cola, A. Deriu, M. Sampoli, and A. Torcini,
J. Chem. Phys. {\bf 104}, 4223 
(1996);
 S.-H. Chen, P. Gallo, F. Sciortino, and P. Tartaglia, Phys. Rev. E
 {\bf 56}, 4231 (1997).

\bibitem{fabbian}
F. Sciortino, L. Fabbian, S. H. Chen, and P. Tartaglia, Phys. Rev. E
{\bf 56}, 5397 (1997);
L. Fabbian, F. Sciortino, and P. Tartaglia, 
J. Non-Cryst. Solids
{\bf 235-237}, 350 (1998).

\bibitem{confined-coupling}
A. Faraone, L. Liu, C. Y. Mou, P. C. Shih, J. R. D. Copley, and S.-H. Chen,
J. Chem. Phys. {\bf 119}, 3963 
(2003).

\bibitem{models-rot}
L. Liu, A. Faraone, and S.-H. Chen,  Phys. Rev. E {\bf 65}, 041506 (2002).

\bibitem{semischematicMCT}
L. Fabbian, F. Sciortino, F. Thiery, and P. Tartaglia, Phys. Rev. E
{\bf 57}, 1485 (1998);
L. Fabbian, R. Schilling, F. Sciortino, P. Tartaglia, and C. Theis,
Phys. Rev. E {\bf 58}, 7272 (1998);
L. Fabbian, F. Sciortino, and P. Tartaglia, 
Phil. Mag. B {\bf 77}, 499 
(1998).

\bibitem{MMCT}
 L. Fabbian, A. Latz, R. Schilling, F. Sciortino, P. Tartaglia, and
 C. Theis, Phys. Rev. E {\bf 62}, 2388 (2000). 

\bibitem{trans-rot-coupl}
A. Faraone, L. Liu, and S.-H. Chen,
J. Chem. Phys. {\bf 119}, 6302 
(2003).

\bibitem{masaki}
J.~Higo, M.~Sasai, H.~Shirai, H.~Nakamura, and T.~Kugimiya,
Proc. Nat. Acad. Sci. {\bf 98}, 5961 (2001).

\bibitem{yeh}
Y.-L. Yhe and C.-Y. Mou, J. Phys. Chem B {\bf 103}, 3699 (1999).

\bibitem{binder}
See, e.g., K. Binder, ed.,
{\it Monte Carlo Methods in Statistical Physics\/}
(Springer-Verlag, Berlin, 1979). 

\bibitem{spce}
 H. J. C. Berendsen, J. R. Grigera, and T. P. Stroatsma,
J. Phys. Chem. {\bf 91}, 6269 (1987).

\bibitem{realH2O} Microwave spectroscopy and infrared spectroscopy on
H$_2$O at equilibrium in the gas phase give 0.9575\AA\ for the O-H
distance and 104.51$^o$ for the HOH angle [D. R. Lide {\it CRC Handbook
of Chemistry and Physics, 84th Edition\/} (CRC Press, Boca Raton FL,
2003)].


\bibitem{berendsen} H. J. C. Berendsen, J. P. M. Postma, W. F. van
Gunsteren, A. DiNola, and J. R. Haak, J. Phys. Chem. {\bf 81}, 3684
(1984).

\bibitem{reactionfield}  
O. Steinhauser, Mol. Phys. {\bf 45}, 335 (1982).


\bibitem{feller-flory} W. Feller, {\it An Introduction to Probability
Theory and Its Applications, Vol.~1, 2nd Edition\/} (John Wiley \& Sons,
New York, 1960), pp.~225--226; P. J. Flory, {\it Statistical Mechanics
of Chain Molecules\/} (John Wiley \& Sons, New York, 1969), p.~16.

\bibitem{note2} In Eq.~(\ref{cgsd}) we normalize the average with the
time-dependent number of molecules $n_i(t)$, instead of the total number
of cells inside the box as in Ref.~\cite{masaki}. Our choice has the
advantage of giving rise to a value of the average dipole in a box
$\vec{d}^{v}_i$ independent of the system density (see for example
\cite{mathias}). However, as a consequence, the distribution of
$\vec{d}^{v}_i$ is not Gaussian, as would be expected by normalizing
by a constant factor. We have verified that by using a constant
normalization factor we recover a Gaussian distribution. Moreover, we
have verified that our final results are not affected by the choice of
the normalization factor in Eq.~(\ref{cgsd}).

\bibitem{mathias}
G. Mathias and P. Tavan, J. Chem. Phys. {\bf 120}, 4393 (2004).

\bibitem{note1} The definition in Eq.~(\ref{cgsd}) differs from the one
introduced in Ref.~\cite{masaki}, where the coarse-grained site-dipole
field is averaged over spheres with radius $R$ and centered at a
distance shorter than $2R$. The definition in Ref.~\cite{masaki}
emphasizes the spatial-patterns of coarse-grained site-dipoles, because
each molecular dipole contributes to the coarse-grained site-dipole for
all the (overlapping) spheres which contain the same molecule. We
therefore expect to find patterns that survive for a time shorter than
that measured in Ref.~\cite{masaki}. Indeed, we do not find strong
evidence of surviving patterns in bulk water within our time resolution.

\bibitem{note3} At $T=300$~K we find a autocorrelation time $\tau_{r}$
approximately 10 times smaller than the persistence time found in
Ref.~\cite{masaki} for the site-dipoles patterns.  This is due to the
difference in the definition in Eq.~(\ref{cgsd}). We verify that, by
adopting the same definition of Ref.~\cite{masaki}, we can reproduce the
bulk-water results of Higo et al. See also \cite{note1}.

\bibitem{MCT}
See for example
W. G\"otze, J. Phys.: Condens. Matter {\bf 11}, A1 
(1999);
W. G\"otze and L. Sjogren, Rep. Prog. Phys. {\bf 55}, 241 
(1992); 
E. Leutheusser,
Phys. Rev. A {\bf 29}, 2765 
(1984);
W. G\"otze, in {\it Liquids, Freezing and the Glass Transition},
Proceedings of the Les Houches Summer School of Theoretical Physics,
Session LI, 1989, edited by J. P. Hansen, D. Levesque, and
J. Zinn-Justin (North-Holland, Amsterdam, 1991);
 A. P. Sokolov, J. Hurst, and D. Quitmann, Phys. Rev. B {\bf 51}, 12~865
 (1995).

\bibitem{Schilling97}
R. Schilling and T. Scheidsteger,
Phys. Rev. E {\bf 56}, 2932  (1997).

\bibitem{Kammerer97}
S. Kammerer, W. Kob, and R. Schilling,
Phys. Rev. E {\bf 56}, 5450 (1997).


\bibitem{orientational}
T. Franosch, M. Fuchs, W. G\"otze, M. R. Mayr, and A. P. Singh,
Phys. Rev. E {\bf 56}, 5659 (1997);
W. G\"otze, A. P. Singh, and T. Voigtmann,
Phys. Rev. E {\bf 61}, 6934 
(2000);
S.-H. Chong and W. G\"otze,
Phys. Rev. E {\bf 65}, 051201 (2002);
Phys. Rev. E {\bf 65}, 041503 (2002).

\end{references}
\end{document}